# CARACTERISATION ELECTROMAGNETIQUE DE MILIEUX HETEROGENES NATURELS - APPLICATION AU SUIVI DE L'HUMIDITE DU SOL PAR RADIOMETRIE MICRO-ONDE

F. Demontoux*, G. Ruffié*, J.P. Wigneron**,MJ Escorihuela***, J. Grant **
\* Laboratoire PIOM-ENSCPB-UMR 5501- 16 av Pey-Berland 33607 Pessac
f.demontoux@enscpb.fr
\*\*INRA-Unité de Bioclimatologie, BP 81, Villenave d'Ornon Cedex 33883
\*\*\*CESBIO 18 avenue. Edouard Belin, bpi 2801 31401 Toulouse cedex 9

## 1. INTRODUCTION

L'humidité de surface du sol est une variable clé pour décrire les échanges d'eau et d'énergie entre la terre et l'atmosphère. En hydrologie et en météorologie, la quantité d'eau contenue dans les couches supérieures du sol (0-5cm de la surface) permet d'évaluer le rapport entre l'évaporation réelle et l'évaporation potentielle au niveau d'un sol nu. Il est aussi possible de déterminer la répartition des précipitations en eaux de ruissellements ou en eaux « stockées » ou d'autres variables comme la conductivité hydraulique. Des études ont montré que les capteurs micro-ondes passifs à 1.4 GHz étaient les mieux adaptés, comme technique de télédétection, pour réaliser un suivi de l'humidité de surface des sols. Une mission de l' ESA (European Space Agency) est en cours de développement pour un lancement prévu en 2007: la mission SMOS (Soil Moisture and Ocean Salinity) ([2] Kerr et al., 2001). Une mission analogue de la NASA est aussi en cours: le projet HYDROS. La thématique principale associée à ces observations est le suivi des échanges hydriques au niveau planétaire dans le cadre du changement climatique. Ce projet européen regroupe des instituts de recherche en France (CESBIO, Météo-France, INRA) et dans de nombreux pays européens (Espagne, Italie, Hollande, Danemark, etc.). Les observations micro-ondes permettent d'estimer la permittivité équivalente de surface des sols, qui est le paramètre principal influant sur la mesure en bande L. Cependant, l'effet de la couverture de végétation, la présence de litière, la température, la rugosité du sol, la couverture neigeuse et la topographie , etc. ont également un effet sur l'émissivité micro-ondes en surface.

Le but des travaux de recherche que nous proposons est l'amélioration de la compréhension des effets de telle structure. En particulier les effets de la litière et des hétérogénéités du sol sont probablement importants mais encore très méconnus ([3]Wigneron et al., 2001). Pour cela et dans un premier temps nous avons mis au point une approche expérimentale en laboratoire, mobile pour aller sur différents terrains. Cela permet d'effectuer des mesures pour des configurations variées en terme instrumental (en fréquence, en temporel, polarisation, incidence, bi-statique, effet Brewster, saut de fréquence …) et en terme de conditions de surface (sol homogène ou hétérogène, présence ou non de litière plus ou moins humide, etc.). L'utilisation de lois d'échelles est aussi envisagée. Des mesures au laboratoire en guide d'onde (figure 1) nous ont permis de caractériser les différents composants de la structure géologique (terre, roches) et de vérifier le modèle de Dobson [4] couramment utilisé.

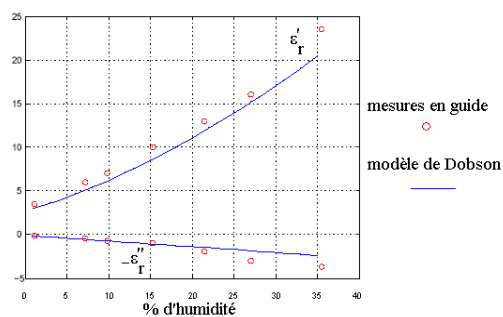

*Figure 1 : Comparaison mesures en guide* WR650 – *modèle de Dobson [4] à 1.4GHz*

## 2. ETUDE DE LA REPONSE ELECTROMAGNETIQUE DE STRUCTURE GEOLOGIQUE

Dans un deuxième temps, nous avons développé des programmes de simulation (méthode des éléments finis) qui permet de simuler les bancs de mesure complets. Le premier modèle utilisé (Figure 2) représente une mesure en réflexion. Des comparaisons ont été effectuées entre des mesures et des simulations (HFSS-Ansoft, méthode des éléments finis) sur des configurations « simples » (figure 2) [6]. Elles montrent qu'il est possible d'obtenir un bon accord entre les résultats expérimentaux et les simulations.. Les figures 2 présentent les résultats du module et de la phase du coefficient de réflexion normalisés par une mesure sur une plaque court circuit. La comparaison des résultats expérimentaux et ceux issus de simulations montre une bonne corrélation.

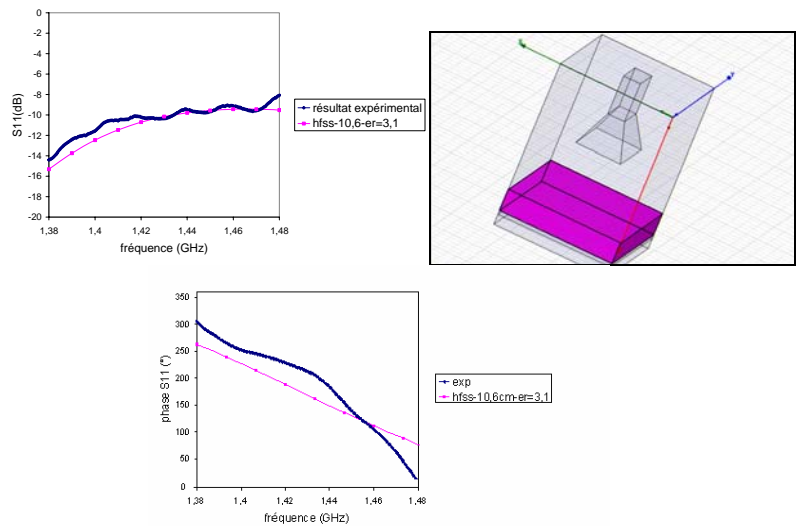

*Figure 2 : résultats de mesures et de calculs en réflexion sur une structure monocouche*

Le deuxième modèle réalisé représente le banc de mesure bi-statique (Figure 3).

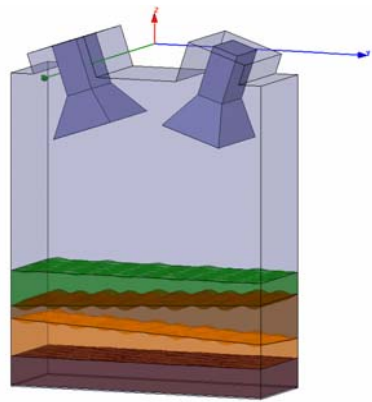

*Figure 3 : Modèle numérique de simulation des mesures bi-statiques*

Ce modèle numérique permet une meilleure interprétation des mesures. Une étape de validation des simulations (par inter-comparaison simulations / observations expérimentales) a été effectué. La Figure 4 présente les résultats obtenus en effectuant les mesures sur de l'eau.

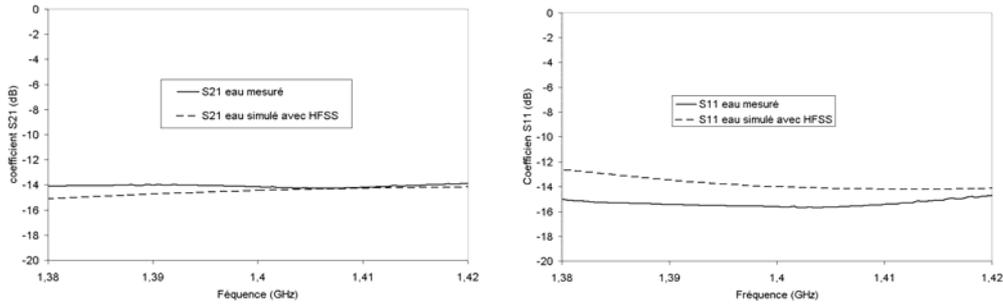

*Figure 4 : comparaison de résultats expérimentaux et numériques d'études bi-statiques*

Le programme permet d'étudier la réponse électromagnétique de milieux géologiques en effectuant des études sur de nombreux paramètres telles que la permittivité, les épaisseurs ou la forme des strates ou encore la rugosité de surface. Nous pouvons aussi effectuer une variation des angles d'inclinaison (angle incidence et polarisation) des deux cornets. Nous comptons créer une base de données simulées et expérimentales sur un très large domaine de configuration de surface. Un modèle FDTD est aussi utilisé [5]. Il permet lui aussi d'introduire de nombreux paramètres tels que la porosité ou la topologie de surface ; la non homogénéité de l'humidité ou de densité de la terre ou encore la présence de cailloux et de strates de différents matériaux

## 3. EMISSIVITE ET HUMIDITE DES SOLS

Les études précédentes nous permettent de définir précisément les relations entre des structures géologiques complexes et les propriétés des ondes réfléchies par ces structures. Le radiomètre, quand à lui mesure la température de brillance du corps. Celle-ci peut être directement rattachée à la température apparente du corps et au coefficient de réflexion de la structure. Ce dernier est très lié à la permittivité des milieux qui elle-même est fonction de l'humidité des sols. Nous avons donc développé une méthode de calcul qui nous permet d'évaluer l'émissivité du sol dont le comportement électromagnétique a été simulé à l'aide de notre modèle. Les valeurs d'émissivité équivalente ainsi obtenues serviront à créer une base de données. Nous souhaitons ainsi améliorer la compréhension de l'influence de paramètres, tels que la rugosité de surface ou les gradients d'humidité, sur les mesures d'émissivité. Les résultats du modèle seront confrontés à la base de données expérimentales acquise sur le site SMOSREX à Toulouse où un radiomètre à 1.4GHz est installé en permanence depuis 2002 (expérience Météo-France/CESBIO/INRA/ONERA) sur une jachère. Sur ce terrain, les effets de l'hétérogénéité du sol et de la litière sont relativement marqué ([1] De Rosnay et al., 2005).

Deux approches peuvent être envisagées pour le calcul analytique de l'émissivité de nos structures. La première est une approche « cohérente » du problème. Elle tient compte de l'amplitude et de la phase du champ réfléchi dans le milieu. La seconde est une approche « non – cohérente ». Dans ce cas ; seul l'amplitude est prise en compte. L'approche « cohérente » est adaptée si les variations de permittivité sont beaucoup plus petites que la longueur d'onde dans le milieu. L'approche « non-cohérente » est destinée davantage aux structures comportant des couches de taille comparable à la longueur d'onde. Les méthodes analytiques utilisées généralement sont limitées car elles ne permettent pas d'introduire des milieux non homogènes ou des rugosités entre les strates par exemple. Dans ce dernier cas une correction des calculs d'émissivité est envisageable mais limitée. Nous avons développé un modèle numérique (méthode des éléments finis) qui permet de calculer la réflectivité équivalente d'une structure géologique en introduisant différents paramètres tel que la rugosité de surface par exemple. Une méthode analytique (logiciel SIMULTIMAT ) a aussi été utilisée afin de valider les premiers résultats de notre modèle numérique (Figure 5). Cette méthode est basée sur le calcul du coefficient de réflexion de nos structures à l'aide d'un produit de matrices représentant les interfaces entre chaque couche ainsi que le déphasage et l'atténuation dans chaque couche.

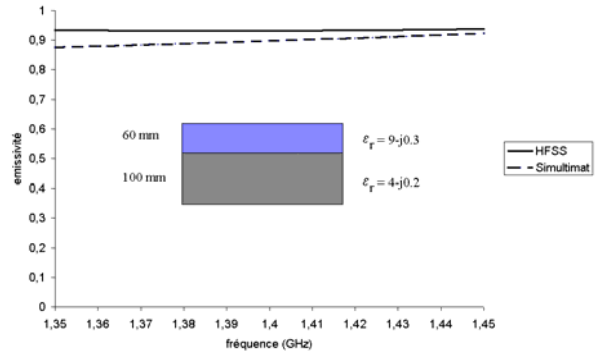

*Figure 5 : comparaison de calculs d'émissivité sur une structure bi-couche*

## 4. CONCLUSION

Nous avons débuté l'exploitation de notre modèle en calculant l'émissivité équivalente d'un sol en fonction de l'humidité (Figure 6). Pour effectuer ces calculs nous avons repris les mesures que nous avons effectuées (Figure 1).

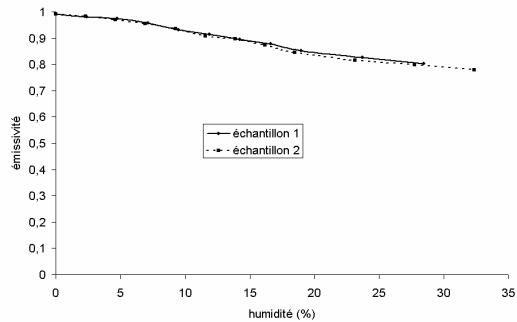

*Figure 6 : suivi de l'émissivité d'un sol en fonction de l'humidité*

Les résultats obtenus avec nos deux méthodes vont être maintenant confrontées aux mesures de terrain. Les structures géologiques étudiées seront plus complexes. Les prochaines études porteront donc sur l'introduction de rugosité de surface ainsi que sur l'effet d'une litière végétale qui peut être très hétérogènes. Enfin nous aborderons l'effet de gradients d'humidité dans le sol ou dans la litière.

[1] De Rosnay P., Y. Kerr, J.-C. Calvet, J.-P. Wigneron, F. Lemaître, M.-J. Escorihuela, J. Munoz Sabater, K. Saleh, N. E. D. Fritz, G. Cherel, R. Durbe, A. Kruszewski, P. Waldteufel, L. Coret, G. Dedieu, 'SMOSREX: A Long Term Field Campaign Experiment for Soil Moisture and Land Surface Processes Remote Sensing, to be submitted, 2005.
[2] Kerr Y. H., P.Waldteufel,J.-P. Wigneron ,J. Font, M.Berger,Soil Moisture Retrieval from Space: The Soil Moisture and Ocean Salinity (SMOS) Mission', IEEE Trans. Geosc. Remote Sens., 39(8) :1729-1735, 2001.
[3] Wigneron J.-P., L. Laguerre, Y. Kerr, 'A simple parameterization of the L-band Microwave Emission from Rough Agricultural Soils', IEEE Trans. Geosc. Remote Sens., 39(8) :1697-1707, 2001.
[4] Mironov, V.L., Dobson, M.C., Kaupp, V.H., Komarov, S.A. Kleshchenko, V.N. Generalized refractive mixing dielectric model for moist soils, IEEE Transactions on Geoscience and Remote Sensing. Volume 42, Issue 4, April 2004, Pages 773-785
[5]E.Heggy., P. Paillou, F. Costard, N. Mangold, G. Ruffie, F. Demontoux, J. Grandjean, J.M. Malezieux « *Local geoelectrical models of the Martian subsurface for shallow groundwater detection using sounding radars* » Journal.Geophys.Research., 2003, vol108n°E4,10.1029/2002JE1871
[6] F. Demontoux, G.Ruffié, J.P. Wigneron, M-J Escorihuela. Amélioration de l'étude de l'humidité de sols par radiométrie. Caractérisation et modélisation diélectriques de profils géologiques.JNM 2005, Nantes, 2005
[7] Microwave remote sensing Volume I, Ulaby, Moore, Fung, Artech House